\documentclass[10pt]{article}

\usepackage{amsmath}
\usepackage{amssymb}

\usepackage{graphicx}

\usepackage{cite}

\usepackage{color} 

\usepackage[labelfont=bf,labelsep=period,justification=raggedright]{caption}

\makeatletter
\renewcommand{\@biblabel}[1]{\quad#1.}
\makeatother

\date{}

\begin{document}

\begin{flushleft}
{\Large
\textbf{Emergence of switch-like behavior in a large family of simple biochemical networks}
}
\\
Dan Siegal-Gaskins$^{1,2,\ast}$, 
Maria Katherine Mejia-Guerra$^{2}$, 
Gregory D. Smith$^{3}$,
Erich Grotewold$^{2}$
\\
\bf{1} Mathematical Biosciences Institute, The Ohio State University, Columbus, OH 43210, USA
\\
\bf{2} Department of Molecular Genetics and Plant Biotechnology Center, The Ohio State University, Columbus, OH 43210, USA
\\
\bf{3} Department of Applied Science, The College of William and Mary, Williamsburg, VA 23187, USA
\\
$\ast$ E-mail: dsg@mbi.osu.edu
\end{flushleft}

\section*{Abstract}
Bistability plays a central role in the gene regulatory networks (GRNs) controlling many essential biological functions, including cellular differentiation and cell cycle control.   However, establishing the network topologies that can exhibit bistability remains a challenge, in part due to the exceedingly large variety of GRNs that exist for even a small number of components.  We begin to address this problem by employing chemical reaction network theory in a comprehensive \emph{in silico} survey to determine the capacity for bistability of more than 40,000 simple networks that can be formed by two transcription factor-coding genes and their associated proteins (assuming only the most elementary biochemical processes).  We find that there exist reaction rate constants leading to bistability in $\sim$90\% of these GRN models, including several circuits that do not contain any of the TF cooperativity commonly associated with bistable systems, and the majority of which could only be identified as bistable through an original subnetwork-based analysis.   A topological sorting of the two-gene family of networks based on the presence or absence of biochemical reactions reveals eleven minimal bistable networks (i.e., bistable networks that do not contain within them a smaller bistable subnetwork).   The large number of previously unknown bistable network topologies suggests that the capacity for switch-like behavior in GRNs arises with relative ease and is not easily lost through network evolution.  To highlight the relevance of the systematic application of CRNT to bistable network identification in real biological systems, we integrated publicly available protein-protein interaction, protein-DNA interaction, and gene expression data from \emph{Saccharomyces cerevisiae}, and identified several GRNs predicted to behave in a bistable fashion.  

\section*{Author Summary}

  
Switch-like behavior is found across a wide range of biological systems, and as a result there is significant interest in identifying the various ways in which biochemical reactions can be combined to yield a switch-like response.   In this work we use a set of mathematical tools from  \emph{chemical reaction network theory} that provide information about the steady-states of a reaction network irrespective of the values of network rate constants, to conduct a large computational study of a family of model networks consisting of only two protein-coding genes.  We find that a large majority of these networks ($\sim$90\%) have (for some set of parameters) the mathematical property known as bistability and can behave in a switch-like manner.  Interestingly, the capacity for switch-like behavior is often maintained as networks increase in size through the introduction of new reactions.   We then demonstrate using published yeast data how theoretical parameter-free surveys such as this one can be used to discover possible switch-like circuits in real biological systems.   Our results highlight the potential usefulness of parameter-free modeling for the characterization of complex networks and to the study of network evolution, and are suggestive of role for it in the development of novel synthetic biological switches.   
  
  


\section*{Introduction}
Bistability---the coexistence of two stable equilibria in a dynamical system---is responsible for the switch-like behavior seen in a wide variety of {cell biological networks, such as those involved in} signal transduction \cite{Pomerening:2008p324}, cell fate specification \cite{Lai:2004p700,Laslo:2006p319,Huang:2007p613}, cell cycle regulation \cite{Cross:2002p881},  apoptosis \cite{Bagci:2006p606,Eissing:2004p607,Legewie:2006p604}, and in regulating extracellular DNA uptake (competence development) \cite{Avery:2005p1014}.    Evidence for bistable networks has been found in experimental observations of the hysteretic (i.e., history dependent) response to stimuli that is commonly associated with bistability \cite{Ninfa:2004p769,Guidi:1997p1008}, for example in the Cdc2 activation circuit in {\em Xenopus} egg extracts \cite{Pomerening:2003p970,Sha:2003p1019} and in the lactose utilization network in {\em E. coli} \cite{Ozbudak:2004p320}.  Complementing experimental analyses,  mathematical tools such as bifurcation theory can be used to determine if a particular network---written as a set of ordinary differential equations (ODEs)---is bistable \cite{Tyson:2001p603}.   However, because the dynamical behavior of a network is dependent on the values of the system parameters (e.g., reaction rates), and the number of parameters required for an accurate description of even simple systems is typically large and uncertain, new bistable circuit architectures tend to be identified only slowly and on a network-by-network basis.

Chemical reaction network theory (CRNT), which gives conditions for the existence, multiplicity, and stability of steady states in systems of nonlinear ODEs derived from mass-action kinetics \cite{Feinberg:1987p913,Feinberg:1988p828,Ellison:2000p845}, offers a novel framework for the rapid identification of network topologies with the capacity for bistability (herein referred to as bistable networks).  Importantly, CRNT is applicable without specific knowledge of the system parameters.   This ability to study network characteristics in a parameter-free context is particularly beneficial in cell and developmental biology, given the high level of uncertainty in parameter values \cite{Kaltenbach:2009p855}.  As a result, CRNT has found a number of biological applications \cite{Conradi:2005p390,OteroMuras:2009p854,Craciun:2006p389,SiegalGaskins:2009p708}.  Still, considering its potential for large-scale analyses, the use of CRNT has been fairly limited. 

Here, we apply CRNT to reaction network models representing a broad class of small gene regulatory networks (GRNs): those consisting of two transcription factor (TF)-coding genes and their associated proteins.   {Our comprehensive parameter-free survey resulted in the identification of 36,771 bistable GRN architectures (out of a total of 40,680)}, including eleven without the TF cooperativity typically associated with switch-like circuits.    Approximately 40\% of the bistable systems were confirmed as such using existing computational tools, with the remainder identified through the novel concept of network ancestry, in which the presence of a bistable subnetwork can under certain conditions be used to establish bistability in a larger network if the condition that the two network structures have an identical stoichiometric subspace is met (see the following section on CRNT basics).  Despite its large size, the entire two-gene bistable network family can be understood as descended from a set of only eleven minimal bistable networks, that is, bistable networks that do not contain within them a smaller bistable subnetwork and, as a consequence, are rendered monostable by the removal of one or more network reactions.  Using experimental protein-protein interaction, protein-DNA interaction, and gene expression data from \emph{Saccharomyces cerevisiae}, we demonstrate how a general theoretical survey of this kind has unique predictive power to identify bistable modules in {organisms that have not been fully explored from a functional genomics perspective}.    Our results are further suggestive of a role for parameter-free modeling in simplifying the study of complex regulatory networks, understanding network evolution, and designing new synthetic biological circuits. 

\section*{Results} 

\subsection*{Two-gene network construction}

As done previously \cite{SiegalGaskins:2009p708}, we assume classical chemical kinetics and specify gene regulatory networks (GRNs) as sets of elementary biochemical reactions. For a network consisting of $N$ transcription factor genes $\mathrm{X}_i$ and associated proteins $\mathrm{P}_i$ $(i=1,\ldots,N)$, the essential reactions are basal protein production ($\mathrm{X}_i \rightarrow \mathrm{X}_i + \mathrm{P}_i$) and degradation ($\mathrm{P}_i \rightarrow \emptyset$).  Networks may also contain protein dimerization reactions ($\mathrm{P}_i + \mathrm{P}_j \rightleftharpoons \mathrm{P}_i\mathrm{P}_j$), binding of both TF monomers and dimers to the gene promoters ($\mathrm{X}_i + \mathrm{P}_j \rightleftharpoons \mathrm{X}_i\mathrm{P}_j$ and $\mathrm{X}_i + \mathrm{P}_j\mathrm{P}_k \rightleftharpoons \mathrm{X}_i\mathrm{P}_j\mathrm{P}_k$), and protein production from a bound gene ($\mathrm{X}_i\mathrm{P}_j \rightarrow \mathrm{X}_i\mathrm{P}_j + \mathrm{P}_i$ and $\mathrm{X}_i\mathrm{P}_j\mathrm{P}_k \rightarrow \mathrm{X}_i\mathrm{P}_j\mathrm{P}_k + \mathrm{P}_i$).  For reactions of this last type, under our parameter-free framework, the rate of protein production from a bound gene is unspecified and thus may be either higher or lower than the basal rate but can not be zero.   For simplicity, we assume that the promoter of each gene may only be bound by a single monomer or dimer species at any given time, or they may remain unoccupied.  We further assume that, while degradation is considered for monomeric TFs, all TF dimers are stable to proteolytic degradation;  the validity of this assumption and its implications are discussed below.  

A variety of networks may be constructed by combining these reactions, subject to certain logical constraints (e.g., the presence of a dimer-promoter binding reaction requires the inclusion of the dimer formation reaction) and with the requirement that every network includes the necessary basal TF production and degradation reactions.  In the two-gene case ($N$=2), there are 4 essential reactions and 23 additional reactions (Table~1) that may be combined to form 40,680 different networks.   The total number of networks is smaller than might be expected (i.e., less than $2^{23}$) as a result of reaction dependencies (Table~1) and network symmetries; for example, the network consisting of reactions {\em k}, {\em q}, and {\em w} is functionally equivalent to the that with reactions {\em i}, {\em l}, and {\em r}, and as a result we do not include the latter and other symmetric networks like it in the total.  

{It should be noted that within this set of two-gene networks there are a small number for which there is no coupling between the two genes.  Given that there are twelve possible one-gene networks for both $\mathrm{X}_1$/$\mathrm{P}_1$ and  $\mathrm{X}_2$/$\mathrm{P}_2$  independently (see \cite{SiegalGaskins:2009p708}), the total number of unique decoupled two-gene networks is 12(12+1)/2=78, the number of distinct pairs of one-gene circuits.  The presence of 78 decoupled two-gene networks was verified by searching through the full list of 40,680 networks for those lacking the basic coupling reactions {\em b}, {\em c}, {\em j}, {\em n}, and {\em o} (Table~1).}

\subsection*{Chemical reaction network theory basics}

Given the centrality of CRNT to our analysis, we provide here a primer on the relevant aspects of the theory and illustrate them with the rudimentary two-gene network that consists of only the essential basal protein production and degradation reactions (Figure~\ref{CRNT}).   

At the heart of the theory is the concept of network {\em complexes}, formally the chemical species or linear combinations of species which occur on either side of a chemical equation.   A reaction network can be visualized as a directed graph where each of these complexes appears only once at the heads and/or tails of reaction arrows.   A collection of complexes connected by arrows is referred to as a {\em linkage class}.   The complexes and linkage classes for our rudimentary network are highlighted in Figure~\ref{CRNT} in yellow and with dashed lines, respectively.  

Every complex in the network can also be represented as a vector in an appropriate vector space; in a network of $N$ species, the complex vectors lie in $\mathbb{R}^N$.  Reactions also have associated vectors (termed {\em reaction vectors}), which are constructed by subtracting the ÔreactantÕ complex vectors from the ÔproductÕ complex vectors.  The size of the largest linearly independent set of reaction vectors is the {\em rank} of the network, and the set of all possible linear combinations of reaction vectors (i.e., their span) is referred to as the {\em stoichiometric subspace} of the network.   This subspace plays an important role in setting boundaries on the system behavior: although the species' concentrations may evolve with time, they are ultimately constrained within surfaces that are parallel translations of the stoichiometric subspace.  Exactly which surface (or {\em stoichiometric compatibility class}) the concentrations are constrained to is defined by the initial conditions.  

For a system with $n$ complexes, $l$ linkage classes, and rank $s$, the network {\em deficiency} $\delta$ is defined as $\delta=n-l-s$. A number of theorems regarding the stability properties of networks are based on the deficiency, including the deficiency zero and deficiency one theorems \cite{Feinberg:1987p913,Feinberg:1988p828}.   

Advanced deficiency theory (ADT) \cite{Ellison:2000p845} is required for networks with a deficiency greater than one.   The ADT algorithm, detailed in \cite{Ellison:1998p904} and implemented in the Chemical Reaction Network Toolbox software package (http://www.chbmeng.ohio-state.edu/$\sim$feinberg/crntwin/), constructs and attempts to solve systems of equalities and inequalities that are based on the network structure.  If no solutions (which together with the equality and inequality systems are known as `signatures' of the reaction network) can be found, then the network does not have the qualitative capacity to support multiple steady states.  However, if signatures can be found, then the network can support multiple steady states, and the Toolbox will produce example rate constants and associated steady states consistent with the mass-action ODE description of the network, as well as report the stability characteristics of the steady states.   It should be emphasized that ADT cannot guarantee bistability even if the network does  support multiple steady states, as they may be unstable.  Nevertheless, with its substantial analytical power and ease of use, ADT has played a role in a number of recent studies \cite{SiegalGaskins:2009p708,Conradi:2007p681,Flockerzi:2008p1025,Miller:2008p1074,SaezRodriguez:2008p784}.

\subsection*{Preliminary bistable network identification}

All of the two-gene networks modeled were found by the Chemical Reaction Network Toolbox (herein referred to as simply the Toolbox) to have a deficiency of two or more, necessitating the use of ADT in their analyses.   Screening the Toolbox-generated ADT analysis reports, we determined that of the 40,680 networks surveyed, 18,352 ($\sim$45\%) have the capacity for multiple steady states, with 14,721 of these being confirmed as bistable with example rate constants (see Materials and Methods for a description of the screening procedure).  Only 2,654 networks ($\sim$6.5\%) cannot be bistable regardless of the parameter values.   For the remaining 19,674 networks, ADT could neither establish nor rule out the capacity for multiple steady states, and as a result we refer to these as `unknown' networks.   It is noteworthy that the fraction of networks of a given size (that is, a given number of reactions) that are unknown increases as the size increases; for example, $>$90\% of networks with more than 21 reactions, and all networks with more than 24 reactions, are unknown (Figure~\ref{unknowns}).   As expected, the stabilities of the decoupled two-gene networks are the same as the constituent one-gene systems previously studied \cite{SiegalGaskins:2009p708}.  

The two smallest bistable networks identified exhibit canonical switch topologies (Figure~\ref{smallest}).  In the double negative feedback circuit shown in Figure~\ref{smallest}A, we find that dimerization of only one of the TFs is sufficient for bistability.  The autoregulatory positive feedback network shown in Figure~\ref{smallest}B is an example of a decoupled two-gene network, with bistability in the concentration of one TF only.   We note that while CRNT does not take into account the strength of the regulation in determining a network's capacity for multiple steady states, the fact that an autoregulatory circuit requires positive feedback in order to achieve bistability is well-established (see, e.g., \cite{Keller:1995p1080,Hasty:2001p1082}).   Bistability via positive autoregulation has also been demonstrated experimentally with synthetic gene circuits in both prokaryotes \cite{Isaacs:2003p1072} and eukaryotes \cite{Becskei:2001p325}. 

\subsection*{Identifying bistability through network ancestry}

The bistable networks shown in Figure~\ref{smallest}, each containing seven reactions, can be `grown' into new eight-reaction networks through the addition of reactions from Table~1:  { reactions {\em a}, {\em b}, {\em d}, {\em g}, {\em i}, {\em j}, {\em q}, or {\em t} to the circuit shown in Figure~\ref{smallest}A, and reactions {\em a}, {\em b},  {\em c}, {\em d}, {\em i}, {\em j}, or {\em n} to the circuit shown in Figure~\ref{smallest}B.  In all cases, the new larger networks were also confirmed by the Toolbox to be bistable.  We may then ask: is bistability}, once established in a `parent' network of $N$ reactions, guaranteed in any `descendant' network of $N+1$ reactions?  ADT alone is not sufficient to answer this question, since {systems were less likely to be characterizable as they increased in size (Figure~\ref{unknowns}).}  However, CRNT does provide a basis for establishing bistability in networks which contain subnetworks known to be bistable:  if following the addition of a reaction the stoichiometric subspace of the descendant network is identical to that of the parent, then the larger network is also bistable for some set of parameter values.  As an intuitive example, one can imagine a situation in which a reaction is added to an existing network, that the surface containing the dynamical trajectories of the network species' concentrations is not changed as a result of the addition, and that the added reaction has only a very small rate constant.   In this case we should not expect a change from whatever qualitative phenomena were there before.  Thus, if the parent network had two stable equilibria, the descendant network will also have two stable equilibria.   Example reactions that do not result in a change in the stoichiometric subspace if added include protein production from a TF-bound gene ($\mathrm{X}_i\mathrm{P}_j \rightarrow \mathrm{X}_i\mathrm{P}_j + \mathrm{P}_i$, since the reaction vectors can be written as linear combinations of the vectors associated with $\mathrm{X}_i + \mathrm{P}_j \rightleftharpoons \mathrm{X}_i\mathrm{P}_j$, $\mathrm{P}_i \rightarrow \emptyset$, and $\mathrm{P}_j \rightarrow \emptyset$).   Beginning with the 14,721 known bistable networks and using this `ancestry'-based method, we identified an additional 22,050 bistable networks.   Some of these networks had been previously found by the Toolbox to have the capacity for multiple steady states, but for which no example parameter sets leading to stable equilibria were given.   The number of networks of each type---bistable by ADT, bistable by ancestry, multiple steady states with unconfirmed stability, monostable, or unknown---are shown as a function of network size in Figure~\ref{network_sizes}.  

\subsection*{Minimal bistable networks}
 
Of the 36,771 bistable systems identified, only eleven do not contain within them a smaller subnetwork that is also bistable.  For these eleven networks, the removal of any single reaction would result in a loss of bistability.    We refer to these networks as {\em minimal bistable networks} (MBNs).   Named according to the reaction labels in Table~1, the MBNs are: {\em kqw}, {\em ckn}, {\em bcdh}, {\em ikno}, {\em jmpsv}, {\em bfjpv}, {\em abejp}, {\em jknptv}, {\em jkmnps}, {\em dhjknp}, and {\em aejknp}.  The two networks shown in Figure~\ref{smallest} are minimal ({\em kqw} and {\em ckn});  the full set is shown in Figure~\ref{minbistables}.   Arrows containing the symbol $(\pm)$ are used in the figure and all that follow to emphasize that, in assessing a networkÕs capacity for multiple steady states, CRNT does not distinguish between up-regulation and down-regulation that results in reduced but non-zero expression.  With the exception of {\em bcdh} (discussed in more detail in the following section), all of these networks contain one or more of the TF dimerization reactions common in bistable GRNs \cite{SiegalGaskins:2009p708}.  It can also be seen that each MBN contains feedback loops that for some parameter sets will be made positive, a characteristic shown to be generally necessary for multiple steady states in a system of ODEs \cite{Cinquin:2002p322}.

\subsection*{Cooperativity-free switches}

Although cooperativity in gene regulation---via either the non-independent binding of TFs to multiple promoter sites or the multimerization of TFs into functional units---is an important component of some bistable networks \cite{Cherry:2000p317,Gardner:2000p768}, it is not necessary for bistability.  Indeed, a number of recent mathematical models of GRNs have shown deterministic bistability without cooperativity of any kind \cite{Francois:2004p314,Lipshtat:2006p857,Buchler:2008p315}.   Among the 40,680 two-gene networks are 45 lacking cooperativity, and of these 31 were found to be monostable, eight were identified as bistable directly by ADT, and three more were identified as bistable by network ancestry.  All of the bistable networks lacking cooperativity can be derived from the MBN {\em bcdh}, which is shown in Figure~\ref{bistable_network} along with a bifurcation diagram showing the existence of two stable equilibria (and an unstable equilibrium) for a range of $\mathrm{P}_1$ degradation rate constants.   The complete set of cooperativity-free bistable networks is shown in Figure~\ref{all_dimer-free_figs}.  An essential feature of these circuits is the competitive binding of $\mathrm{P}_1$ and $\mathrm{P}_2$ to the $\mathrm{X}_2$ promoter.  {Similar competitive or sequestration-type processes have been found to be key components in some switch-like systems \cite{Francois:2004p314,Lipshtat:2006p857,Buchler:2008p315,Sedlak:1995p887,Basak:2008p886}.  

\subsection*{Two-gene networks in \emph{S. cerevisiae}}

To investigate how an \emph{in silico} network topology survey such as this can be used to better understand experimental results, we searched for real biological examples of the bistable networks identified in this study in the model organism \emph{S. cerevisiae}.   To our knowledge, there is no single database that contains \emph{S. cerevisiae} GRN architecture, thus we combined protein-protein and protein-DNA interaction data with gene expression data to establish the large-scale empirical network shown in Supplementary Figure S1.   Included in this network are 148 TFs participating in 205 protein-protein interactions (61 heterodimerization and 144 homodimerization reactions), along with 1,249 interactions between 139 TFs and 208 genes (37 `self-binding' and 1,212 `cross-binding' reactions).  To establish which of the two-gene bistable circuits are present in the yeast network, it was first necessary to `translate' the bistable models from their ideal, theoretical description (that distinguishes between and allows for each elementary reaction) into a format that is more amenable to experimental data mining;  see Supplementary Text S1.   We were then able to identify in the yeast data a total of 1,289 two-gene GRNs, twelve of which have topologies consistent with members of the MBN set (Table 2).   Examples of these are highlighted in the next section.

\section*{Discussion}

The idea of studying theoretical network models generated via `random wiring' was suggested at least fifty years ago by Monod and Jacob \cite{Monod:1961p937}.   Only recently, with the development of powerful computational tools, have a variety of simple gene regulatory and metabolic network topologies been studied with surveys over large ranges of parameter space \cite{Ramakrishnan:2008p687,Ma:2009p674}.   Parameter-free techniques such as CRNT are particularly well-suited for general surveys aimed at bistable network discovery, as they may more definitively answer questions regarding a mass action system's ability to support multiple steady states.   For example, using only the advanced deficiency theory (ADT) algorithm implemented in the Chemical Reaction Network Toolbox we were able to establish that $\sim$36\% of the 40,680 possible unique two-gene networks are bistable for at least some sets of network parameters, another $\sim$9\% have the capacity for multiple steady states (which may or may not be stable), and only $\sim$6.5\% are monostable regardless of the network parameters. 

As network size and complexity increases, the ability of ADT to draw conclusions becomes limited (Figure~\ref{unknowns}).  One method put forward as a way to extend the usefulness of CRNT to larger networks involves the analysis of simpler subnetworks corresponding with elementary flux modes of the system \cite{Conradi:2007p681}.  We have introduced a complementary subnetwork analysis method for identifying bistability, termed network ancestry, which requires only a topological sorting of the networks based on the presence or absence of individual reactions followed by inspection of the network reaction vectors.   If the parent network is determined to be bistable, and if the reaction vectors of the bistable parent and unknown descendant have the same span (i.e., the networks have an identical stoichiometric subspace), then the descendant is also bistable.  As a result of network ancestry, we were able to identify an additional 22,050 networks with previously unknown stability as bistable, $\sim$54\% of the total (Figure~\ref{network_sizes}).  We emphasize that a change in the size of the stoichiometric subspace does not in and of itself imply that bistability will be lost; however, from a purely topological perspective, it may not be obvious what the effect of the change may be.  Our network ancestry method may thus be considered a relatively conservative one for establishing bistability in larger networks.

The assumption of mass action kinetics is an important aspect of CRNT.  Consequently, Michaelis-Menten and Hill-type expressions are not used in our CRN approach, as they require approximations to mass action that cannot be validated in a parameter-free context.  In addition, it was recently demonstrated for a generic two-protein interaction network that bistability present under the `inconsistent' assumption of Michaelis-Menten kinetics is lost when the system is `unpacked' into its fundamental chemical steps \cite{SabouriGhomi:2008p312}.   For our two-gene networks, the Michaelis-Menten and CRN descriptions could be approximately equivalent only for specific parameters, and only if those parameters were such that 1) the DNA-binding reactions reach their equilibria much more quickly than other reactions in the network, and 2) the equilibrium concentrations of any dimer species were proportional to the product of their constituent monomer concentrations \cite{Bundschuh:2003p832}.

In addition to the inherent consistency of CRN models, the mathematical theory applicable to deterministic CRNs offers significant computational advantages over other methods, in particular stochastic simulation.  Furthermore, many deterministically bistable networks have been shown to retain two long-lived states when their models are reformulated to take stochasticity into account \cite{SabouriGhomi:2008p312,Stamatakis:2009p877,Lipshtat:2006p857,Kepler:2001p617}.    Still, as biochemical noise has been shown to drive some systems to exhibit switch-like behavior not predicted by deterministic models \cite{Lipshtat:2006p857,Kepler:2001p617,Blake:2003p1073,Samoilov:2005p871,Arkin:1998p437}, it should be considered in any complete study of a specific network of interest.  For models already formulated as CRNs, stochastic simulation is relatively straightforward (see, e.g., \cite{SabouriGhomi:2008p312,Gillespie:2007p874}).  

We attempted to capture the most prevalent and basic biochemical processes involved in transcriptional regulation in our network model construction, but our formalism is by no means exhaustive.  One mechanism not included and through which networks can achieve the nonlinearity required for bistability is the direct degradation of TF dimers ($\mathrm{P}_i\mathrm{P}_j \rightarrow \emptyset$) \cite{Buchler:2008p315}.  Given that dimerization regularly protects against proteolysis (see, e.g., \cite{Jenal:2003fk, Johnson:1998uq}), its exclusion from our reaction set is reasonable.  Furthermore, for most of the networks analyzed here, the addition of a dimer degradation reaction would have no effect on their capacity for bistability:  since the reaction vector for $\mathrm{P}_i\mathrm{P}_j \rightarrow \emptyset$ can be written as a linear combination of the vectors associated with reactions $\mathrm{P}_i + \mathrm{P}_j \rightleftharpoons \mathrm{P}_i\mathrm{P}_j$, $\mathrm{P}_i \rightarrow \emptyset$ and $\mathrm{P}_j \rightarrow \emptyset$, any descendant network grown from a bistable parent via the addition of a dimer degradation reaction would have the same stoichiometric subspace and would be bistable as a result of network ancestry.   

There remains a large amount of additional biological detail which could be incorporated in future surveys, including post-translational modification, multiple promoter binding sites, and the location of regulatory elements relative to the genes (which has been shown to play a role in network bistability \cite{Kelemen:2010p891}).  However, any increase in the level of detail would result in an increase in the combinatorial complexity and the size of the survey.  For example, whereas the set of one-gene networks are constructed using combinations of 5 different reactions \cite{SiegalGaskins:2009p708}, and our two-gene networks using 23 reactions (Table~1), the addition of a third gene alone would lead to 60 different reactions that could be `wired' together.   With the current version of the Toolbox taking (at best) many seconds to import, analyze, and export the results for every network model, it is perhaps not an ideal software package for surveys significantly larger than this one.  New software implementations of CRNT continue to be developed (e.g., \cite{Soranzo:2009p706}), and we anticipate that future programs will allow for even more comprehensive computational studies.   In the meantime, network ancestry offers an attractive solution to the problem of scalability and applicability of CRNT to more complex networks:  once all fundamental chemical reactions involved in any network of interest are identified, one could assemble the minimal network topologies covering all possible unique stoichiometric subspaces and probe that smaller set of networks for bistability.    In essence, network ancestry allows for the reduction of the problem of determining a large network's qualitative capacity for bistability to one of identifying the minimal bistable subnetworks within it. 

There is a strong biological motivation to consider individual networks as parents and descendants with a topological ordering:  rather than appearing \textit{de novo}, modern GRNs grow from ancestor network kernels through mechanisms such as gene duplication and the accretion of protein domains \cite{Chothia:2003p961,Force:2005p962,Lynch:2003p963,Buljan:2010p971}.   Domain accretion, for example the acquisition of a DNA-binding domain by a monomer (modeled in this work by the addition of one of the promoter binding reactions $a$, $b$, $c$, or $d$), has been proposed to be particularly important for eukaryotic evolution \cite{Babushok:2007p973,Koonin:2000p921}.   And there is evidence suggesting an even more direct role for bistability in evolution:  it is the primary requirement for epigenetic inheritance mechanisms known to have important evolutionary effects \cite{Veening:2008p1013,Jablonka:1998p539}, and can also lead to increased population fitness in stressful or changing environments \cite{Bishop:2007p1085,Kussell:2005p1086} by driving an increase in phenotypic heterogeneity \cite{Dubnau:2006p1084}.    Thus, the eleven MBNs identified here (Figure~\ref{minbistables}), which differ from monostable networks by just a single reaction, may represent an interesting class of networks from the standpoint of evolutionary {biology, as it may be that} similarly-minimal networks have played an important role in functional development and/or speciation.  

We used the results of our \textit{in silico} analysis to motivate a search of---and add functional context to---existing yeast protein-DNA and protein-protein interaction data, and in doing so were able to identify a number of two-gene systems with topologies consistent with {bistability.  For example, the} FKH1 and FKH2 genes (and their associated proteins Fkh1p and Fkh2p, which compete for target promoter occupancy \cite{Hollenhorst:2001p952}) compose a network with a topology similar to the MBN {\em bcdh} (Table 2).   FKH1 and FKH2 belong to the pervasive winged-helix/forkhead (FOX) family of TFs and are essential for proper regulation of the yeast cell cycle \cite{Zhu:2000p956}.  Other FOX genes have previously been shown to be involved in important biological functions including cell cycle regulation and cell differentiation \cite{Hannenhalli:2009p955}, two processes for which GRN bistability has been implicated \cite{Lai:2004p700,Laslo:2006p319,Huang:2007p613, Cross:2002p881}.  

Additional gene pairs of interest include NRG1/RIM101 and OAF1/PIP2, which are components of GRNs with topologies similar to that of MBNs {\em abejp} and {\em aejknp}, respectively.  The Rim101p and Nrg1p proteins, both identified previously as transcriptional repressors, are components in an extracellular pH-responsive differentiation pathway in yeast \cite{Lamb:2003p957}.   Further evidence suggestive of bistability in this system can be found in {\em C. albicans}, in which Rim101p and Nrg1p homologs regulate the morphological switch \cite{Bensen:2004p953} associated with a dramatic change in the pathogen's virulence \cite{Kumamoto:2005p958}.   Oaf1p and Pip2p, on the other hand, are involved in the production of peroxisomal proteins in the presence of fatty acids \cite{Baumgartner:1999p951}, and have been shown to be involved in the coordination of two different transcriptional responses to oleate \cite{Smith:2007p959}.  We emphasize that while the two-gene networks identified through our analysis are not guaranteed to be bistable, their known topologies and functions make them excellent bistable network candidates, providing powerful hypotheses for further experimentation.    The same approach may be used to provide guidance or functional context to any system for which the necessary interaction data is available.

High-throughput parameter-free analysis holds potential, not just as a tool for the study of natural systems, but also as a design aid in the growing field of synthetic biology \cite{Andrianantoandro:2006p106,Ellis:2009p914}.  For example, a survey such as this can provide inspiration for the development of new bistable switches and a library of models to draw from;  already we have proposed a set of novel bistable networks that lack cooperativity and which may be particularly good designs as a result (e.g., because they do not require any `extra' engineering of dimerization domains).     At the very least, such a broad application of CRNT may be used to rule out (possibly large numbers of) designs incapable of bistability.  CRNT can be similarly used to rule out circuits without the capacity for sustained oscillations \cite{Feinberg:1987p913} or those which cannot exhibit `absolute concentration robustness'  \cite{Shinar:2010p897}.    

It is worth emphasizing that the region of parameter space supporting bistability in any individual network cannot be determined via parameter-free techniques alone.  For example, it may be that the necessary parameter values lie outside the range of biological reality or are difficult to engineer, or that the size of the bistable region of parameter space is exceedingly small.    However, in many large-scale studies, such as those that resulted in the yeast data sets used in this work, a high degree of biochemical detail is simply nonexistent.   While this lack of quantitative detail can make some analyses of biological networks challenging, it also opens up opportunities for parameter-free studies to provide experimental guidance and new functional insights \cite{Bailey:2001p850}.    Once identified, potentially interesting network architectures may be analyzed in more detail, with rate constants chosen, for example, by Monte Carlo sampling of parameter space.

\section*{Materials and Methods}
\subsection*{Two-gene network construction}

Two-gene networks were generated in MATLAB (2009a, The MathWorks, Inc.) by first enumerating all possibilities and then removing one network from each symmetric pair (defined by two functionally-equivalent networks which can be made identical through a simple change of component subscripts).   The heterodimers $\mathrm{P}_1\mathrm{P}_2$ and $\mathrm{P}_2\mathrm{P}_1$ were assumed to be equivalent.  

\subsection*{Chemical reaction network theory analysis and network screening}   
  
Advanced deficiency theory analysis was done using a preliminary version of the Chemical Reaction Network Toolbox v2.0 (http://www.chbmeng.ohio-state.edu/$\sim$feinberg/crntwin/) made available to us by M.~Feinberg and automated with AutoIt v3 (http://www.autoitscript.com/autoit3/index.shtml).  

Networks were screened based on the content of the analysis reports generated by the Toolbox. These reports, though unique to each network, all contain one of three statements:  either the network ``DOES have the capacity for multiple steady states", ``CANNOT admit multiple positive steady states", or ``MAY have the capacity for multiple steady states".    Networks with reports containing one of the latter two statements were labeled monostable and unknown, respectively.   If a network was determined by ADT to have the capacity for multiple steady states, the analysis report also contained one (or more) example set(s) of rate constants and the associated pair(s) of distinct steady states.  However, each steady state may be either asymptotically stable, unstable, or with a stability that is ``left undetermined".   Only those networks that could support multiple steady states and for which an example pair of asymptotically stable steady states was given were deemed to be bistable networks.  This is not to imply that multiple steady state networks without such an example are not bistable, only that we were unable to confirm their bistability with ADT.  The screening procedure is shown schematically in Figure~\ref{network_sorting}.

Network ancestry and minimal bistable network analysis was done using MATLAB.  Parent and descendant network pairs were found by simple comparison of the networksÕ stoichiometric subspaces and their constituent reactions (descendant networks contain all the same reactions as their parents plus one additional reaction).   Cooperativity-free networks were identified by their lack of dimerization reactions, since by construction, the model genes do not have two TF binding sites that could be occupied simultaneously and there are no multi-protein complexes larger than dimers.          

Additional data analysis was done with MATLAB and Mathematica (Wolfram Research, Inc.).   The bifurcation plot shown in Figure~\ref{bistable_network} was generated using XPPAUT (http://www.math.pitt.edu/$\sim$bard/\newline xpp/xpp.html).

\subsection*{Identification of bistable networks in \emph{S. cerevisiae}} 

A set of 228 yeast genes previously established as coding for transcriptional regulators \cite{Harbison:2004fk,Drobna:2008rt} was used as the primary source for candidate network TF genes (Supplementary Table S1).  Protein-protein interactions were retrieved from the BioGRID database \cite{Breitkreutz:2008yq} (Supplementary Table S2) and protein-DNA interactions were retrieved from the Yeastract database \cite{Teixeira:2006fk} (Supplementary Table S3).   The effect of the protein-DNA interactions on target gene expression (activation or repression) is usually unknown, and any information suggestive of a particular effect was used supplementarily in the network discovery process (Supplementary Table S4).

\section*{Acknowledgments}
We are grateful to M.~Feinberg for the preliminary version of the Chemical Reaction Network Toolbox v2.0 and for many useful discussions, and to G.\ Craciun, J.\ Stelling, A.\ P.\ Arkin, J.\ Paulsson, and J.\ J.\ Collins for their comments as well.  DSG was jointly mentored by GDS and EG.


\bibliography{2GeneNetBibliography}

\clearpage

\section*{Figure Legends}
\begin{figure}[!ht]
\begin{center}
\includegraphics[width=3.27in]{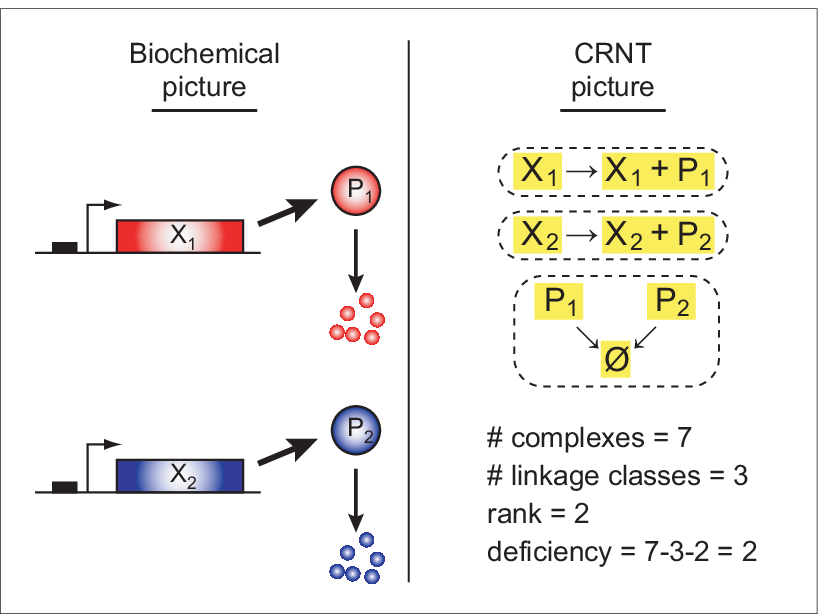}
\end{center}
\caption{
{\bf Rudimentary two-gene network consisting of only basal protein production and degradation.}  In the `CRNT picture', complexes are highlighted in yellow and linkage classes are identified with dashed lines.  Labeling scheme is adopted from  \cite{Shinar:2010p897}.
}
\label{CRNT}
\end{figure}

\begin{figure}[!ht]
\begin{center}
\includegraphics[width=3.27in]{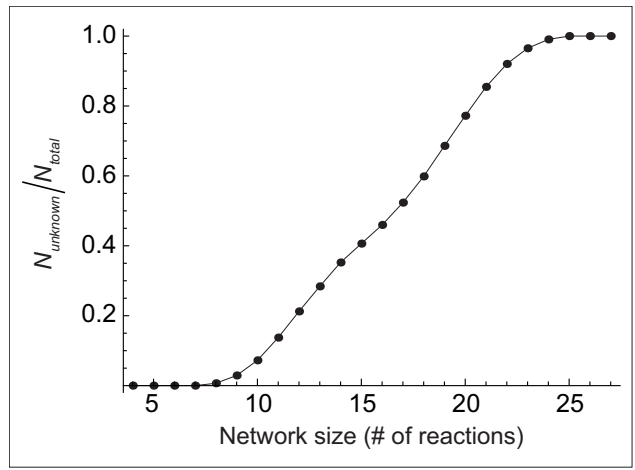}
\end{center}
\caption{
{\bf Fraction of networks which cannot have their stability established by advanced deficiency theory (ADT), as a function of network size.} 
}
\label{unknowns}
\end{figure}

\begin{figure}[!ht]
\begin{center}
\includegraphics[width=3.27in]{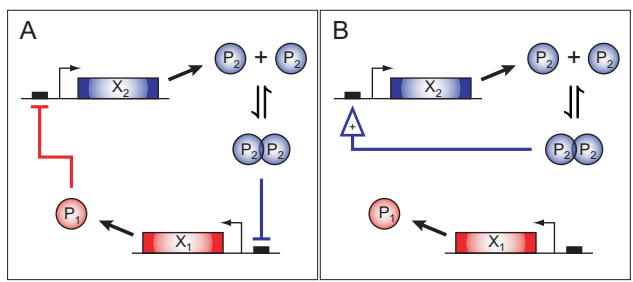}
\end{center}
\caption{
{\bf The smallest two-gene bistable networks found with ADT.}  (A)  A double negative feedback circuit, in which dimerization of only one of the TFs is sufficient for bistability.  (B)  An autoregulatory positive feedback circuit.  The two genes are uncoupled and the bistability is in the concentration of one TF only.  In both (A) and (B), degradation of the TF monomers is not shown.  
}
\label{smallest}
\end{figure}

\begin{figure}[!ht]
\begin{center}
\includegraphics[width=3.27in]{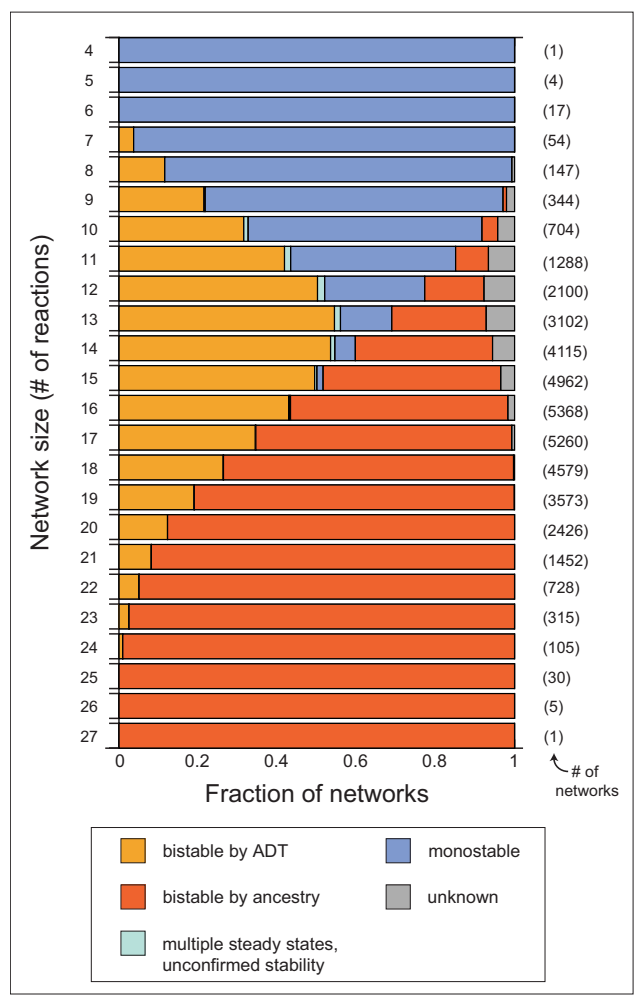}
\end{center}
\caption{
{\bf The fraction of networks of each size that were established as bistable by ADT, bistable by network ancestry, having multiple steady states with unconfirmed stability, monostable, or with an unknown capacity for multiple stable steady states.}  Network size is determined only by the number of reactions (from Table~1) that are present.  The total number of networks of each size is shown in parentheses.
}
\label{network_sizes}
\end{figure}

\begin{figure}[!ht]
\begin{center}
\includegraphics[width=4.86in]{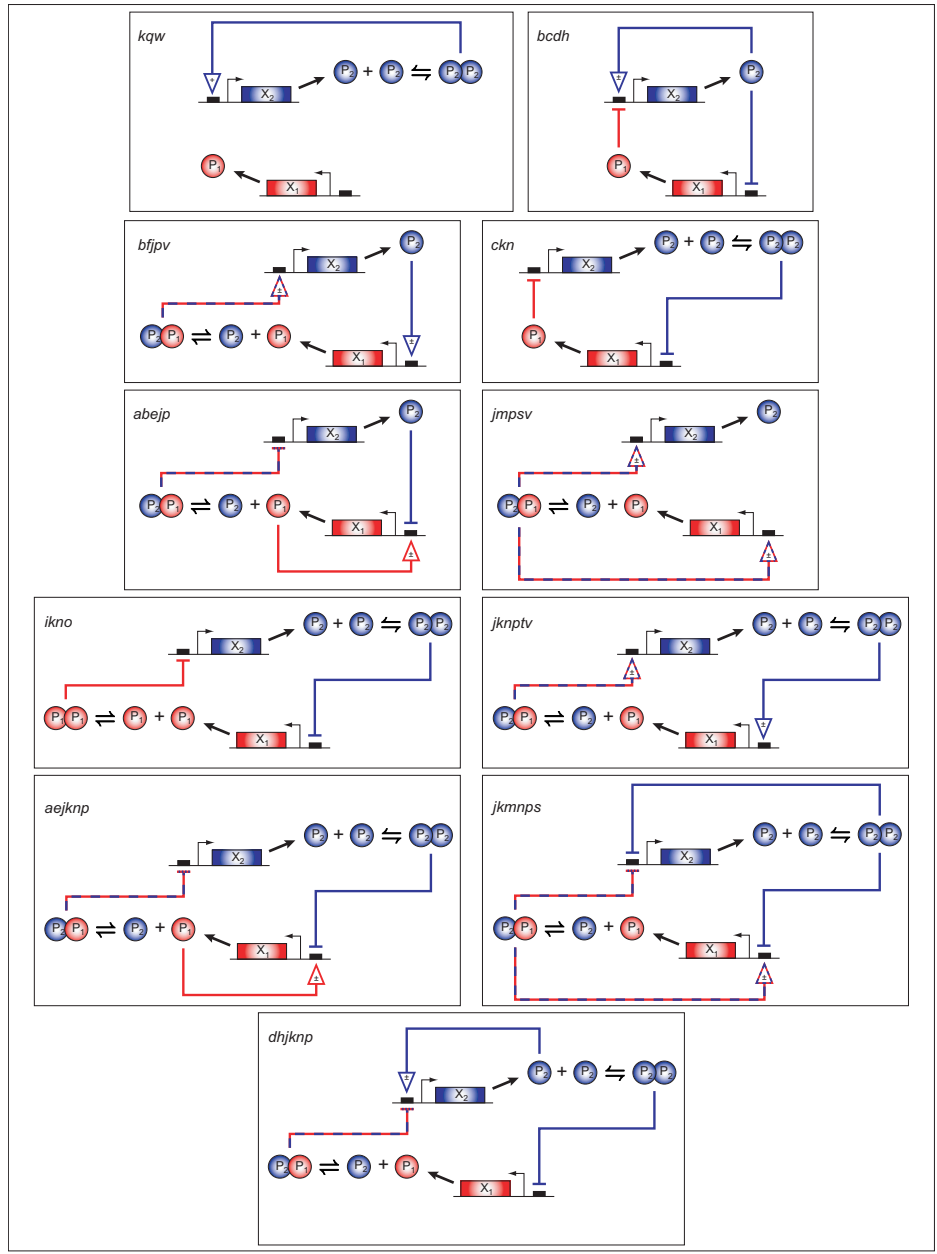}
\end{center}
\caption{
{\bf Minimal bistable networks.}   Only 11 of the 36,771 bistable networks identified lose bistability by the removal of any network reaction. That is, only 11 of the bistable networks contain no subset of reactions which is also bistable.  Dashed-and-colored lines indicate regulation by heterodimer.  Horizontal bars represent purely-repressive TF binding, and arrows indicate TF production from a bound gene (at a non-zero rate that may be either higher or lower than the basal rate). Degradation of the TF monomers is not shown.
}
\label{minbistables}
\end{figure}

\begin{figure}[!ht]
\begin{center}
\includegraphics[width=3.27in]{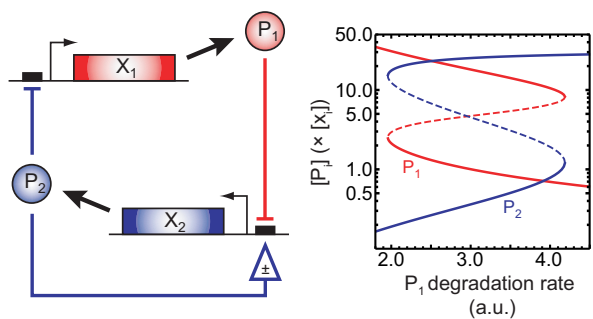}
\end{center}
\caption{
{\bf Example of a bistable network lacking cooperativity.}  TF $\mathrm{P}_2$ plays a dual role as an activator of $\mathrm{X}_2$ and a repressor of $\mathrm{X}_1$.  The bifurcation plot shows the stable (solid lines) and unstable (dashed lines) steady state protein concentrations, in units relative to the DNA concentration, for one set of parameter values as a function of the $\mathrm{P}_1$ degradation rate.  The network ODEs and parameter values are given in the Supplementary Text S1.
}
\label{bistable_network}
\end{figure}

\begin{figure}[!ht]
\begin{center}
\includegraphics[width=4.86in]{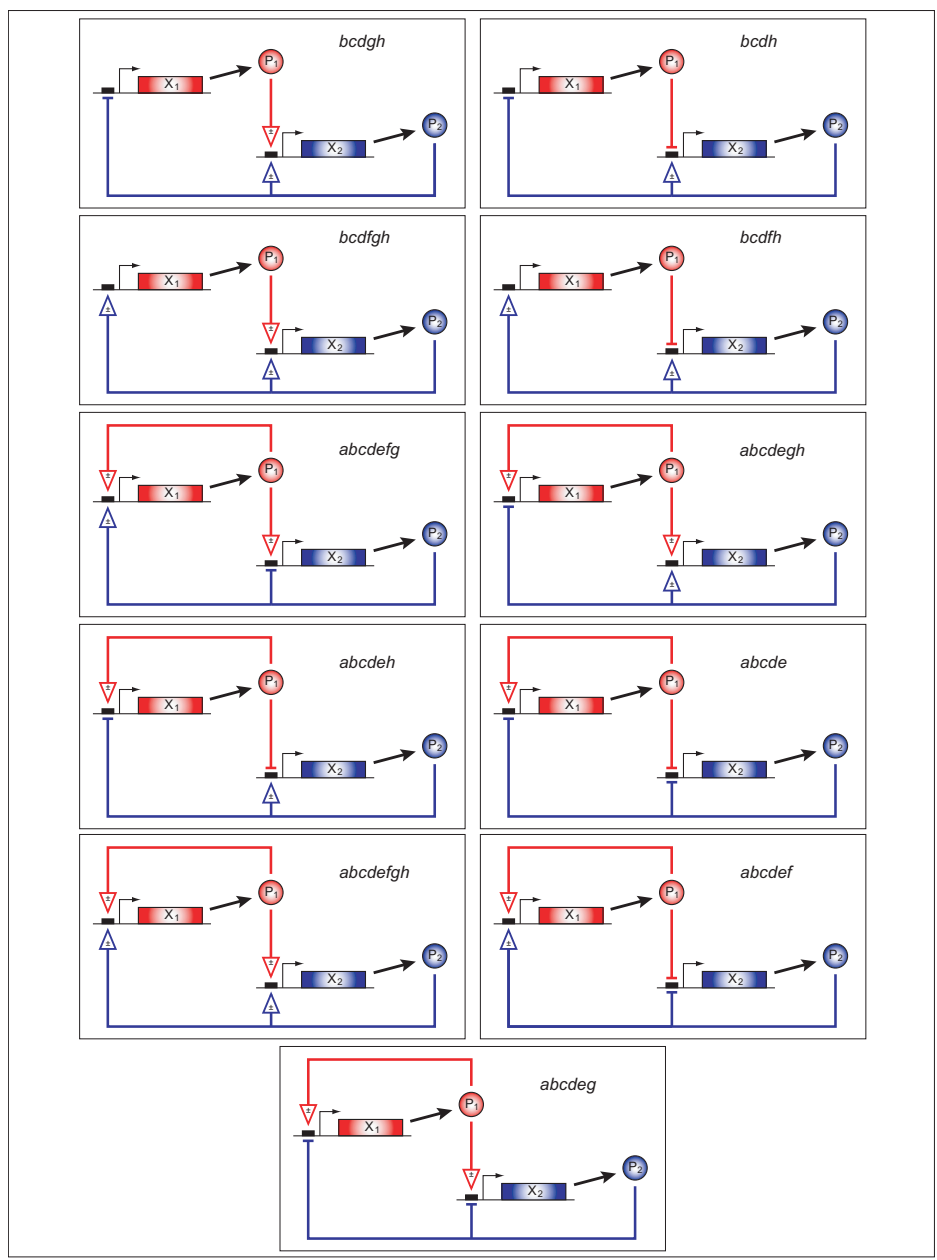}
\end{center}
\caption{
{\bf Bistable networks without cooperativity.}   Of the 45 two-gene networks lacking dimerization, 11 were identified as bistable either directly by advanced deficiency theory analysis or via network ancestry.  All the dimer-free bistable networks shown here can be derived from the minimal bistable network \emph{bcdh} through the addition of reactions from Table 1.  Horizontal bars represent purely-repressive TF binding, and arrows indicate TF production from a bound gene (at a non-zero rate that may be either higher or lower than the basal rate).  Degradation of the TF monomers is not shown.
}
\label{all_dimer-free_figs}
\end{figure}

\begin{figure}[!ht]
\begin{center}
\includegraphics[width=4.86in]{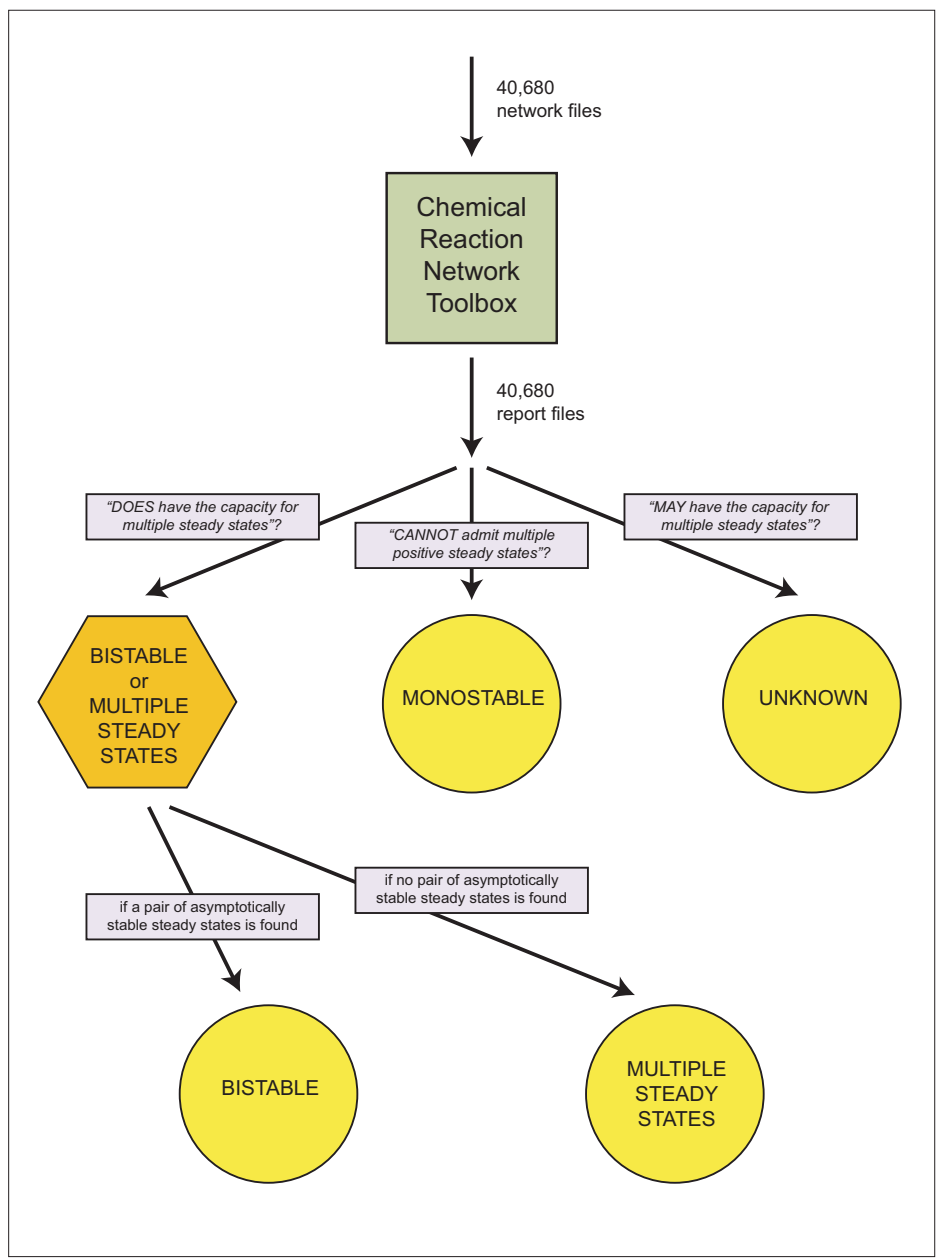}
\end{center}
\caption{
{\bf Screening networks for different steady state behaviors.}   Networks are initially screened by the content of analysis reports produced by the Chemical Reaction Network Toolbox.    The networks designated `multiple steady states' are those determined by ADT to have the capacity for multiple steady states but for which no example pair of asymptotically stable steady states could be found by the program.   Bistable networks are those for which an example pair of asymptotically stable steady states was reported.  The complete sorting procedure is described in Materials and Methods.
}
\label{network_sorting}
\end{figure}

\clearpage

\section*{Supplementary Text, Figure, and Table Legends}

\noindent {\bf Supplementary Text S1. ODEs and parameter values for Fig. 6, and the method used in translating bistable network models into the experimental data mining format.} \\

\noindent {\bf Supplementary Table S1.  List of genes/proteins considered as transcriptional regulators in yeast}. Data taken from \cite{Harbison:2004fk,Drobna:2008rt}.\\

\noindent {\bf Supplementary Table S2. List of protein-protein interactions.} Physical protein-protein interactions between yeast transcriptional regulators extracted from BioGRID database \cite{Breitkreutz:2008yq}.\\

\noindent {\bf Supplementary Table S3. List of protein-DNA interactions.} Physical protein-DNA interactions were extracted from Yeastract database \cite{Teixeira:2006fk}.\\

\noindent {\bf Supplementary Table S4. Transcriptional effect of protein-DNA interactions.}\\

\noindent {\bf Supplementary Figure S1.  Large-scale GRN in \emph{S. cerevisiae}.}  GRN was generated through the combination of protein-protein interaction, protein-DNA interaction, and gene expression data.\\

\clearpage

\section*{Tables}
\begin{table}[!ht]
\caption{
\bf{Reactions combined to generate the 40,680 unique networks of two genes and two gene products}}
\begin{tabular}{clcl}
Reaction & & & \cr 
label $r_i$ & Reaction & Dependencies & Biochemical process \cr 
\hline
\noalign{\vskip 2pt}
$^\ast$ & $\mathrm{X}_1 \rightarrow \mathrm{X}_1 + \mathrm{P}_1$  & -- & gene $\mathrm{X}_1$ basal protein production \cr 
$^\ast$ & $\mathrm{X}_2 \rightarrow \mathrm{X}_2 + \mathrm{P}_2$  & -- & gene $\mathrm{X}_2$ basal protein production\cr 
$^\ast$ & $\mathrm{P}_1 \rightarrow \emptyset$  & -- & protein $\mathrm{P}_1$ degradation\cr 
$^\ast$ & $\mathrm{P}_2 \rightarrow \emptyset$  & -- & protein $\mathrm{P}_2$ degradation\cr 
{\em a} & $\mathrm{X}_1 + \mathrm{P}_1 \rightleftharpoons \mathrm{X}_1\mathrm{P}_1$  & -- & binding of $\mathrm{P}_1$ to the $\mathrm{X}_1$ promoter \cr 
{\em b} & $\mathrm{X}_1 + \mathrm{P}_2 \rightleftharpoons \mathrm{X}_1\mathrm{P}_2$  & -- & binding of $\mathrm{P}_2$ to the $\mathrm{X}_1$ promoter \cr  
{\em c} & $\mathrm{X}_2 + \mathrm{P}_1 \rightleftharpoons \mathrm{X}_2\mathrm{P}_1$  & -- & binding of $\mathrm{P}_1$ to the $\mathrm{X}_2$ promoter \cr  
{\em d} & $\mathrm{X}_2 + \mathrm{P}_2 \rightleftharpoons \mathrm{X}_2\mathrm{P}_2$  & -- & binding of $\mathrm{P}_2$ to the $\mathrm{X}_2$ promoter \cr
{\em e} & $\mathrm{X}_1\mathrm{P}_1 \rightarrow \mathrm{X}_1\mathrm{P}_1 + \mathrm{P}_1$  & {\em a} & production of $\mathrm{P}_1$ from a $\mathrm{P}_1$-bound gene \cr 
{\em f} & $\mathrm{X}_1\mathrm{P}_2 \rightarrow \mathrm{X}_1\mathrm{P}_2 + \mathrm{P}_1$  & {\em b} & production of $\mathrm{P}_1$ from a $\mathrm{P}_2$-bound gene \cr 
{\em g} & $\mathrm{X}_2\mathrm{P}_1 \rightarrow \mathrm{X}_2\mathrm{P}_1 + \mathrm{P}_2$  & {\em c} & production of $\mathrm{P}_2$ from a $\mathrm{P}_1$-bound gene \cr  
{\em h} & $\mathrm{X}_2\mathrm{P}_2 \rightarrow \mathrm{X}_2\mathrm{P}_2 + \mathrm{P}_2$  & {\em d} & production of $\mathrm{P}_2$ from a $\mathrm{P}_2$-bound gene \cr  
{\em i} & $\mathrm{P}_1 + \mathrm{P}_1 \rightleftharpoons \mathrm{P}_1\mathrm{P}_1$  & -- & homodimerization of $\mathrm{P}_1$ \cr 
{\em j} & $\mathrm{P}_1 + \mathrm{P}_2 \rightleftharpoons \mathrm{P}_1\mathrm{P}_2$  & -- & heterodimerization of $\mathrm{P}_1$ and $\mathrm{P}_2$ \cr 
{\em k} & $\mathrm{P}_2 + \mathrm{P}_2 \rightleftharpoons \mathrm{P}_2\mathrm{P}_2$  & -- & homodimerization of $\mathrm{P}_2$ \cr  
{\em l} & $\mathrm{X}_1 + \mathrm{P}_1\mathrm{P}_1 \rightleftharpoons \mathrm{X}_1\mathrm{P}_1\mathrm{P}_1$ & {\em i} & binding of $\mathrm{P}_1\mathrm{P}_1$ dimer to the $\mathrm{X}_1$ promoter  \cr 
{\em m} & $\mathrm{X}_1 + \mathrm{P}_1\mathrm{P}_2 \rightleftharpoons \mathrm{X}_1\mathrm{P}_1\mathrm{P}_2$ & {\em j} & binding of $\mathrm{P}_1\mathrm{P}_2$ dimer to the $\mathrm{X}_1$ promoter \cr 
{\em n} & $\mathrm{X}_1 + \mathrm{P}_2\mathrm{P}_2 \rightleftharpoons \mathrm{X}_1\mathrm{P}_2\mathrm{P}_2$ & {\em k} & binding of $\mathrm{P}_2\mathrm{P}_2$ dimer to the $\mathrm{X}_1$ promoter \cr 
{\em o} & $\mathrm{X}_2 + \mathrm{P}_1\mathrm{P}_1 \rightleftharpoons \mathrm{X}_2\mathrm{P}_1\mathrm{P}_1$  & {\em i} & binding of $\mathrm{P}_1\mathrm{P}_1$ dimer to the $\mathrm{X}_2$ promoter  \cr 
{\em p} & $\mathrm{X}_2 + \mathrm{P}_1\mathrm{P}_2 \rightleftharpoons \mathrm{X}_2\mathrm{P}_1\mathrm{P}_2$  & {\em j} & binding of $\mathrm{P}_1\mathrm{P}_2$ dimer to the $\mathrm{X}_2$ promoter \cr 
{\em q} & $\mathrm{X}_2 + \mathrm{P}_2\mathrm{P}_2 \rightleftharpoons \mathrm{X}_2\mathrm{P}_2\mathrm{P}_2$  & {\em k} & binding of $\mathrm{P}_2\mathrm{P}_2$ dimer to the $\mathrm{X}_2$ promoter \cr 
{\em r} & $\mathrm{X}_1\mathrm{P}_1\mathrm{P}_1 \rightarrow \mathrm{X}_1\mathrm{P}_1\mathrm{P}_1 + \mathrm{P}_1$ & {\em i, l} & production of $\mathrm{P}_1$ from a $\mathrm{P}_1\mathrm{P}_1$-bound gene \cr 
{\em s} & $\mathrm{X}_1\mathrm{P}_1\mathrm{P}_2 \rightarrow \mathrm{X}_1\mathrm{P}_1\mathrm{P}_2 + \mathrm{P}_1$ & {\em j, m}  & production of $\mathrm{P}_1$ from a $\mathrm{P}_1\mathrm{P}_2$-bound gene \cr 
{\em t} & $\mathrm{X}_1\mathrm{P}_2\mathrm{P}_2 \rightarrow \mathrm{X}_1\mathrm{P}_2\mathrm{P}_2 + \mathrm{P}_1$ & {\em k, n} & production of $\mathrm{P}_1$ from a $\mathrm{P}_2\mathrm{P}_2$-bound gene \cr 
{\em u} & $\mathrm{X}_2\mathrm{P}_1\mathrm{P}_1 \rightarrow \mathrm{X}_2\mathrm{P}_1\mathrm{P}_1 + \mathrm{P}_2$ & {\em i, o}  & production of $\mathrm{P}_2$ from a $\mathrm{P}_1\mathrm{P}_1$-bound gene \cr 
{\em v} & $\mathrm{X}_2\mathrm{P}_1\mathrm{P}_2 \rightarrow \mathrm{X}_2\mathrm{P}_1\mathrm{P}_2 + \mathrm{P}_2$ & {\em j, p}  & production of $\mathrm{P}_2$ from a $\mathrm{P}_1\mathrm{P}_2$-bound gene \cr 
{\em w} & $\mathrm{X}_2\mathrm{P}_2\mathrm{P}_2 \rightarrow \mathrm{X}_2\mathrm{P}_2\mathrm{P}_2 + \mathrm{P}_2$ & {\em k, q} & production of $\mathrm{P}_2$ from a $\mathrm{P}_2\mathrm{P}_2$-bound gene \cr 
\hline
\noalign{\vskip 2pt}
\end{tabular}
\begin{flushleft} $^\ast$ These reactions occur in every network.
\end{flushleft}
\label{tab:label1}
\end{table}


\begin{table}[!ht]
\caption{
\bf{Two-gene networks found in \emph{S. cerevisiae} that have topologies consistent with members of the minimal bistable network set}} 
\begin{tabular}{ccc} 
Bistable model$^\ast$ 	& $\mathrm{X}_1$ 	& $\mathrm{X}_2$	\cr 
\hline 
\noalign{\vskip 2pt}					
{\em ckn}		& PDR1 (YGL013C) 	& RPN4 (YDL020C) \cr
{\em bcdh}	& FHL1 (YPR104C)	& MSN4 (YKL062W)	 \cr
{\em bcdh}	& HMS1 (YOR032C)	& YAP6 (YDR259C)	 \cr
{\em bcdh}	& IXR1 (YKL032C)	& PHD1 (YKL043W)	 \cr
{\em bcdh}	& RPN4 (YDL020C)	& YAP1 (YML007W)	 \cr
{\em bcdh}	& FKH1 (YIL131C)	& FKH2 (YNL068C)	 \cr
{\em jknptv}	& MTH1 (YDR277C)	& RGT1 (YKL038W)	 \cr
{\em aejknp}	& OAF1 (YAL051W)	& PIP2 (YOR363C)	 \cr
{\em abejp}	& NRG1 (YDR043C)	& RIM101 (YHL027W) \cr
{\em abejp}	& IFH1 (YLR223C)	& RAP1 (YNL216W)	 \cr
{\em bfjpv}		& KSS1 (YGR040W)	& CST6 (YIL036W)	 \cr
{\em bfjpv}		& OPI1 (YHL020C)	& INO2 (YDR123C)	 \cr
\hline 
\noalign{\vskip 2pt}
\end{tabular}
\begin{flushleft} $^\ast$ Model names refer to the constituent reactions as labeled in Table 1.
\end{flushleft}
\label{tab:label2}
\end{table}

\end{document}